\documentclass[aps,prl,twocolumn,groupedaddress,amsmath,superscriptaddress]{revtex4-2}
\usepackage{caption}
\usepackage{ragged2e}

\usepackage{amsmath}
\usepackage{amssymb}
\usepackage{graphicx} 
\usepackage{subfigure}
\usepackage{overpic}
\usepackage[usenames, dvipsnames]{color}
\usepackage[colorlinks=true,citecolor=blue,urlcolor  = purple]{hyperref}
\usepackage{multirow}
\usepackage{booktabs}
\usepackage{natbib}
\setcitestyle{open={},close={}}

\newcommand{\upcite}[1]{\textsuperscript{\textsuperscript{\cite{#1}}}}

\usepackage{xcolor}
\definecolor{ngreen}{rgb}{0.2,0.7,0.2}

\definecolor{nyellow}{rgb}{1.0,0.75,0.0}

\usepackage{titlesec}
\titleformat*{\section}{\bfseries\raggedright}
\titleformat*{\subsection}{\bfseries\raggedright}

\usepackage{eso-pic} 
\usepackage{lipsum}  

\begin{document}
	
	\title{Exceptional-Point-Induced Nonequilibrium Entanglement Dynamics in Bosonic Networks}

    	\author{
		Chenghe Yu$^{1,\,2,\,*}$,
		Mingsheng Tian$^{1,\,*,\,\S}$,
		Ningxin Kong$^{1}$,
		Matteo Fadel$^{3}$,
		Xinyao Huang$^{4,\,\dagger}$,
		Qiongyi He$^{1,\,2,\,5,\,6,\,\ddagger}$\\
		\bigskip
		\textit{\small$^{1}$State Key Laboratory for Mesoscopic Physics, School of Physics, Frontiers Science Center for Nano-optoelectronics, \\
			\small$\&$ Collaborative Innovation Center of Quantum Matter, Peking University, Beijing 100871, China.\\
			\small$^{2}$Hefei National Laboratory, Hefei 230088, China.\\
			\small$^{3}$Department of Physics, ETH Z\"{u}rich, 8093 Z\"{u}rich, Switzerland\\
			\small$^{4}$School of Physics, Beihang University, Beijing 100191, China\\
			\small$^{5}$Collaborative Innovation Center of Extreme Optics, Shanxi University, Taiyuan, Shanxi 030006, China\\
			\small$^{6}$Peking University Yangtze Delta Institute of Optoelectronics, Nantong 226010, Jiangsu, China}
	}
    

	\begin{abstract}
		Exceptional points (EPs), arising in non-Hermitian systems, have garnered significant attention in recent years, enabling advancements in sensing, wave manipulation, and mode selectivity. However, their role in quantum systems, particularly in influencing quantum correlations, remains underexplored. In this work, we investigate how EPs control multimode entanglement in bosonic chains. Using a Bogoliubov–de Gennes (BdG) framework to describe the Heisenberg equations, we identify EPs of varying orders and uncover spectral transitions between purely real, purely imaginary, and mixed eigenvalue spectra. These spectral regions, divided by EPs, correspond to three distinct entanglement dynamics: oscillatory, exponential, and hybrid. Remarkably, we demonstrate that higher-order EPs, realized by non-integer-$\pi$ hopping phases or nonuniform interaction strengths, significantly enhance the degree of multimode entanglement compared to second-order EPs. Our findings provide a pathway to leveraging EPs for entanglement control and exhibit the potential of non-Hermitian physics in advancing quantum technologies.
		
	\end{abstract}
	
	\maketitle
	
    	\AddToShipoutPictureFG*{%
		\put(50,5){
			\parbox[b][2cm][t]{0.9\textwidth}{\raggedright%
				\footnotesize
				\hrule height 0.4pt width 3cm\relax
				\vspace{0.3cm} 
				$^{*}$ These authors contributed equally to this work.\\
				$^{\S}$ Current address: Department of Physics, The Pennsylvania State University, University Park, Pennsylvania, 16802, USA\\
				$^{\dagger}$ \href{mailto:xinyaohuang@buaa.edu.cn}{xinyaohuang@buaa.edu.cn}\\
				$^{\ddagger}$ \href{mailto:qiongyihe@pku.edu.cn}{qiongyihe@pku.edu.cn}%
			}%
		}%
	}
	
	\section*{Introduction}
	Non-Hermitian physics has garnered significant attention for its ability to capture dissipative and open-system dynamics beyond traditional quantum mechanics\upcite{Ashida2020,El-Ganainy2018}. 
	Among its intriguing phenomena, EPs\upcite{Ozdemir2019,Miri2019,Bergholtz2021}
	---special parameter regimes where eigenvalues and eigenvectors coalesce---offer profound implications in a range of applications, from enhancing sensitivity in sensors\upcite{Chen2017,Yu2020,Ruan2024} 
	to controlling light-matter interactions\upcite{Feng2017,Miao2016,Li2023a}. 
	While EPs are well-studied in classical systems\upcite{Hodaei2014,Doppler2016,ZhangXL2019, He2023, Lee2025}, 
	the investigation of their quantum counterparts is still at an initial stage. 
	Using methods such as extended Hilbert spaces\upcite{Wu2019} and quantum trajectories\upcite{Naghiloo2019} to construct effective non-Hermitian Hamiltonians, or using the Liouvillian superoperator formalism within the Lindblad master equation framework\upcite{Minganti2019,Arkhipov2021}, which explicitly incorporates quantum jump effects, recent work has realized EPs in various quantum systems such as optical systems\upcite{Xiao2020,Ozturk2021,Gao2025}, superconducting circuits\upcite{Naghiloo2019,Chen2021,Chen2022,Abo2024}, ion traps\upcite{Ding2021,Bu2023,Chen2025}, nitrogen-vacancy centers\upcite{Wu2019,Liu2021,Wu2024} and ultracold atoms\upcite{Li2019,Zhao2025}. 
	The coexistence of the non-Hermiticity and quantum properties makes them promising platforms for establishing the relationship between quantum correlations and EPs in few-body systems\upcite{Han2023,Li2023,Teixeira2023,Tang2024}. 

	Extending to the investigation of the broader implications of non-Hermitian physics in multimode systems, previous research has revealed intriguing phenomena in classical regime, such as exceptional topology\upcite{Bergholtz2021}, enhanced sensing via higher-order EPs\upcite{Hodaei2017,Xiao2019}, and the dynamic behavior of first-order moments\upcite{Metelmann2015,Porras2019,Wanjura2020,Tian2023,Gomez-Leon2023,Wanjura2021}. 
	However, the investigation of second-order moments, which are crucial for understanding quantum correlations\upcite{Simon2000,Adesso2007}, remains largely unexplored.
	Instead of addressing the quantum jump effects in dissipative quantum systems\upcite{Naghiloo2019,Minganti2019,Arkhipov2021}, the bosonic chains with mode-hopping and squeezing interactions\upcite{McDonald2018,Mcdonald2020,Luo2022,Busnaina2024,Lee2024,Liu2024,Shi2025,wakefield2024} 
	provide an efficient route to construct non-Hermitian dynamics without dissipation in Hermitian quantum systems. 
	With advances in optomechanics\upcite{Xu2016,Xu2019,Del2022,Wanjura2023,Slim2024} and mechanical systems\upcite{Bild2023,vonLupke24,Marti2024,Yang2024}, the implementation of the desired bosonic chains offers a timely opportunity to explore multimode quantum correlations in the non-Hermitian framework.
	This leads to a fundamental question: Can non-Hermitian physics, especially EPs, help us uncover deeper insights into multimode quantum correlations in an $N$-mode bosonic network?
	
	In this work, we explore the connection between non-Hermiticity and entanglement in bosonic chains with mode-hopping and squeezing interactions. Using the BdG framework to describe the equations of motion, we identify the conditions for second-order and higher-order EPs and analyze their effects on quantum entanglement dynamics. Importantly, we find that EPs divide the parameter space into three distinct regions, characterized by purely imaginary, purely real, and mixed eigenvalue spectra, along with entanglement exhibiting exponential, oscillatory, or hybrid behaviors. Furthermore, we demonstrate that higher-order EPs can significantly enhance the degree of multimode entanglement when parameter are optimally tuned. 
	These results deepen our understanding of the connections between non-Hermitian spectral properties and entanglement dynamics in bosonic systems, offering a practical foundation for utilizing EPs to control quantum correlations in complex dynamical processes.

	\section*{Results}
	\subsection*{Our Model}
\begin{figure}
		\centering
		\includegraphics[width=8cm]{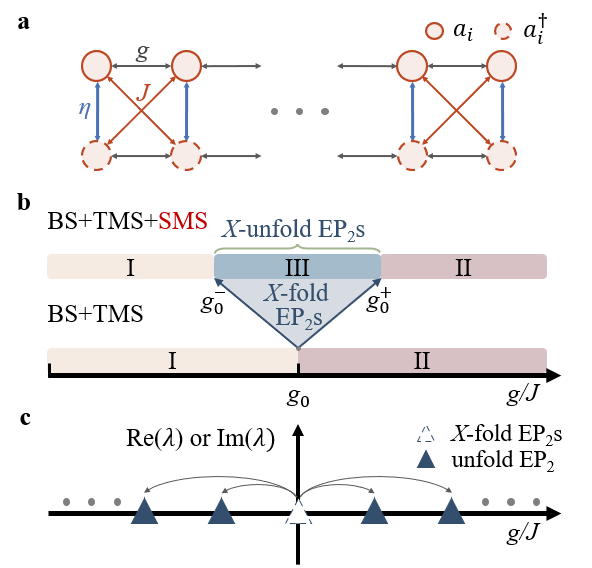}
		\caption{\textbf{Schematic of $N$-mode bosonic chain and the dynamical spectrum divided by EPs.}
			\textbf{a} Schematic of an $ N $-mode bosonic chain with nearest-neighbor mode-hopping and squeezing terms.
			The system is represented graphically using the BdG formalism, which includes BS ($g$), TMS ($J $), and SMS  ($ \eta $) interactions (Supplementary Note 1). \textbf{b-c} Illustration of the dynamical spectrum ($ M $) change affected by the EPs. For a system with only BS and TMS terms, the spectrum is divided into two regions by $ X $-fold EP$_2$s appearing at $g=g_{0}$. In contrast, the introduction of the SMS term splits the $ X $-fold EP$_2$s into $X$-unfold EP$_2$s located in the region of $ g_0^- \leq g \leq g_0^+ $, corresponding to three different regions of the spectrum. Here $X=N$ for even $N$ and $X=N-1$ for odd $N$.} 
		\label{fig1}
	\end{figure}

	\noindent
	We consider a network of $ N $ bosonic modes with nearest-neighbor hopping and squeezing terms, as illustrated in Fig.~\ref{fig1}(a). The Hamiltonian takes the form ($\hbar=1$) 
	\begin{equation}\label{eq1}
		\hat{H}
		= \sum_{j=1}^{N} \frac{\eta}{2}\hat{a}_j^2 + \sum_{j=1}^{N-1} \left( g \hat{a}_j^\dagger \hat{a}_{j+1} + J \hat{a}_j^\dagger \hat{a}_{j+1}^\dagger \right) + \mathrm{h.c.},
	\end{equation}  
	where $\hat{a}_j$ and $\hat{a}_j^\dagger$ denote the mode annihilation and creation operators. Here, $g$ ($g_j=g$), $J$ ($J_j=J$), and $\eta$ ($\eta_j=\eta$) represent the uniform beam-splitter (BS),  two-mode squeezing (TMS), and single-mode squeezing (SMS) rates, respectively.  
	The presence of the squeezing terms (either TMS or SMS) significantly changes the system’s nonequilibrium behaviors. In addition to modifying the dynamics, leading to non-reciprocal and non-Hermitian behavior\upcite{Gomez-Leon2023,Slim2024,Del2022,Wanjura2023}, the squeezing terms are also well-known for generating entanglement. This raises the question:  
	What is the connection between the non-Hermiticity of the dynamic matrix and entanglement? 
	
	To address it, we consider the dynamic matrix $ M $ describing the nonequilibrium behavior of the system through the Heisenberg equation of motion $i(d/dt)\Phi(t)=M\Phi(t)$. Using the vector $\Phi = [\pmb{a} , \pmb{a}^\dagger]^\mathrm{T}$ with $\pmb{a}=[\hat{a}_1,\hat{a}_2,\ldots,\hat{a}_N]^{\mathrm{T}}$, the dynamic matrix $ M $ takes the form of a bosonic BdG matrix  
	\begin{equation}
		M(g, \eta, J)=
		\begin{bmatrix}
			\mathcal{A}(g) & \mathcal{B}(\eta, J) \\
			-\mathcal{B}^\ast(\eta, J) & -\mathcal{A}^\ast(g)
		\end{bmatrix},
		\label{eq:M}
	\end{equation}  
	where $\mathcal{A}(g)$ and $\mathcal{B}(\eta, J)$ are $ N \times N $ parameter matrices (see detailed expressions in Supplementary Note 1). 
	Crucially, the presence of the squeezing terms ensures $\mathcal{B}(\eta, J) \neq 0$, which induces non-Hermitian dynamics without dissipation even though the Hamiltonian itself remains Hermitian\upcite{Ashida2020,Del2022,Wanjura2023}. This approach fundamentally differs from both the construction of effective non-Hermitian Hamiltonians\upcite{Wu2019,Naghiloo2019} and the use of the Liouvillian superoperator formalism\upcite{Minganti2019,Arkhipov2021}, as it generates effective dynamical non-Hermiticity through quantum coherence and parametric driving under BdG framework.

	By transforming the non-Hermitian matrix $ M $ into its Jordan normal form, the structure of EPs in the dynamical spectrum can be identified through its block structure (see details in Supplementary Note 2). For the minimal system size $ N = 2 $, the eigenvalues of the conventional bosonic Kitaev
	chain (BKC) with nearest-neighbor hopping and paring interaction ($g, J\neq0$ and $\eta = 0$) are $\lambda_{1,2} =  \sqrt{g^2 - J^2}$ and $\lambda_{3,4} = - \lambda_{1,2}$. These eigenvalues reveal two degenerate second-order EPs (2-fold EP$_2$s) at $ g = g_0 = J $, dividing the spectrum into two distinct regions: purely imaginary eigenvalues for $ g < g_0 $ and purely real eigenvalues for $ g > g_0 $. This degeneracy of 2-fold EP$_2$s can be broken by adding the SMS term.
	As shown in Fig.~\ref{fig1}(b-c), the 2-fold EP$_2$s are split into two separate EP$_2$s at $ g = g_0^- = |J - \eta| $ and $ g = g_0^+ = |J + \eta| $, respectively. This splitting creates three distinct regions: (I) a purely imaginary region ($ g < g_0^- $), (II) a purely real region ($ g > g_0^+ $), and (III) an intermediate region ($ g_0^- < g < g_0^+ $), where the eigenvalues are mixed, with two being real and the other two being imaginary [Fig.~\ref{fig2}(a-b)].  
	
	Extending to the multimode BKC with $N$ being even,  $ N $-fold EP$_2$s are observed at $ g = g_0 $, separating the purely imaginary region (I, $ g < g_0 $) from the purely real region (II, $ g > g_0 $). The addition of SMS term breaks the degeneracy at $ g = g_0 $, splitting the $ N $-fold EP$_2$s into $ N $-unfold EP$_2$s. For example, in the case of $ N = 4 $, four non-degenerate EP$_2$s can be found, satisfying $ g_0^-=g_1 < g_2 < g_3 < g_4=g_0^+ $. The spectrum in this case exhibits three regions: (I) the purely imaginary region ($ g < g_0^- $), (II) the purely real region ($ g > g_0^+ $), and (III) an intermediate region ($ g_0^- < g < g_0^+ $), where eigenvalues mix real and imaginary parts.  
	For systems with $ N $ being odd, $ (N - 1) $-fold EP$_2$s are found. Unlike the case with even $ N $, odd-number-mode chains do not feature a purely real spectrum after adding SMS term. Instead, it contains imaginary eigenvalues in all regions, resulting in scalable entanglement over time (see details in Supplementary Note 2). 

	\subsection*{Nonequilibrium Entanglement Dynamics Divided by EPs}
	\label{sec3} 
	\noindent
	Multimode systems can present complex entanglement structures, whose characterization poses significant challenges. We focus here on bipartite entanglement, namely entanglement between two partitions of the system, such as between one mode and the rest of the system, which we denote as $(1|N-1)$.
	The entanglement is characterized by the covariance matrix (CM), which describes second-order moments and serves as a common criterion of bipartite entanglement for Gaussian systems\upcite{Simon2000, Adesso2007}.  
	The CM $\sigma$ of an $N$-mode state $\hat{\rho}$ is expressed as a $2N \times 2N$ real symmetric matrix with elements  
	$ 
	\sigma_{ij} = \langle \hat{\beta}_i \hat{\beta}_j + \hat{\beta}_j \hat{\beta}_i \rangle - 2\langle \hat{\beta}_i \rangle \langle \hat{\beta}_j \rangle,  
	$ 
	where $\pmb{\beta} = (\hat{X}_1, \hat{P}_1, \dots, \hat{X}_N, \hat{P}_N)^{\top}$. The operators $\hat{X}_i = (\hat{a} + \hat{a}^\dagger)/\sqrt{2}$ and $\hat{P}_i = -i(\hat{a} - \hat{a}^\dagger)/\sqrt{2}$ represent the amplitude and phase quadratures of the $i$th mode, respectively.   
	Quantification of bipartite entanglement can be achieved through the minimal symplectic eigenvalue, denoted as $\nu_-$, of the CM after performing a partial transpose. Specifically, $\nu_- < 1$ signifies the presence of entanglement, with smaller values of $\nu_-$ indicating stronger entanglement.
	This allows us to track the time evolution of  $\nu_-(t)$ in different parameter regions.

\begin{figure}
		\centering
		\includegraphics[width=8cm]{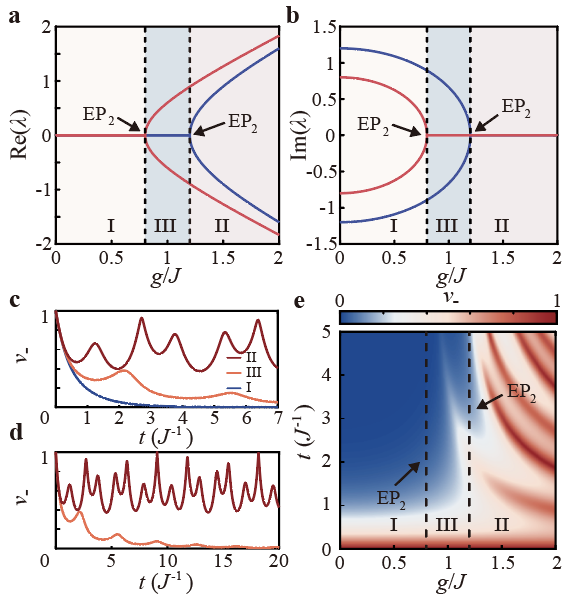}
		\caption{\textbf{Exceptional points and entanglement dynamics for the two-mode system.} The real part \textbf{a} and imaginary part \textbf{b} of four eigenvalues of $M$ versus $g/J$ with $\eta/J = 0.2$. The spectrum is divided into three regions by two EP$_2$s at $g_0^-/J = 0.8$ and $g_0^+/J = 1.2$. \textbf{c} The time evolution of $\nu_-$ for $g/J = 0.79$ (blue), $g/J = 1.19$ (orange) and $g/J = 1.59$ (red), showing three distinct types of the entanglement dynamics, i.e. exponential behavior in region I (blue), oscillatory behavior in region II (red) and mixed behavior in region III (orange). \textbf{d} The long time evolution of $\nu_-$, to show more difference between region II (red) and III (orange). Other parameters are the same as \textbf{c}. \textbf{e} 
			The values of $\nu_-$ versus $g/J$ and $t/J$ with $\eta/J = 0.2$. The representative parameters ($g/J = 0.79$ and $g/J = 1.19$) near the EP$_2$s in \textbf{c} are chosen to illustrate the contrasting entanglement dynamics across these regions, while $g/J = 1.59$ is just chosen to ensure a consistent parametric spacing.
		}
		\label{fig2}
	\end{figure}

	For the two-mode model without SMS interaction,   
	\begin{equation}\label{gjnu}
		\nu_-(t)=\sqrt{\xi(t)-\sqrt{\xi(t)^2-\mathrm{det}\tilde{\sigma}(t)}},
	\end{equation}
	where $\xi(t)=(g^2-J^2\cos(4ct))/c^2,\quad c=\lambda_{1,2}=\sqrt{g^2-J^2}$ and $\tilde{\sigma}(t)$ is the partial transposed CM (PCM), as can be seen in Methods. 
	The time evolution of the parameter $\nu_-$ shows different behaviors in the region of purely imaginary eigenvalues ($g/J < 1$) and purely real eigenvalues ($g/J > 1$), corresponding to a change from exponentially decaying to oscillatory.  
	When adding SMS interaction, the entanglement dynamics become more complex while maintaining close connections to the spectral bifurcations.  In the purely imaginary region (region I) as shown in Fig.~\ref{fig2} (a-b),  the entanglement witness $\nu_-$ exhibits an exponential decay, indicating a monotonic increase in entanglement over time [blue line in Fig.~\ref{fig2}(c)]. In contrast, in the purely real region (region II), entanglement exhibits oscillatory behavior over time [red line in Fig.~\ref{fig2}(c-d)], markedly different from the behavior in region I. In the intermediate region (region III), where the eigenvalues are a mixture of real and imaginary parts, the entanglement dynamics combine exponential-like and oscillatory behaviors [orange line in Fig.~\ref{fig2}(c-d)].  
	We classify these dynamics into three distinct types: type-I, type-II, and type-III [Fig.~\ref{fig2}(e)], corresponding to regions I, II, and III, respectively. Each type exhibits unique nonequilibrium behaviors in both the time and frequency domains (see details in Supplementary Note 3). 
	Extending the analysis to a multimode chain and considering the translation invariance of the bulk modes, a similar transition in bipartite entanglement dynamics induced by EP$_2$s can be observed. 
	These entanglement transitions align well with the spectral bifurcation points, as detailed in Supplementary Note 3. 

	\subsection*{Enhanced entanglement by higher-order EPs}
	\label{sec4}  
	
	\begin{figure}
		\centering
		\includegraphics[width=8cm]{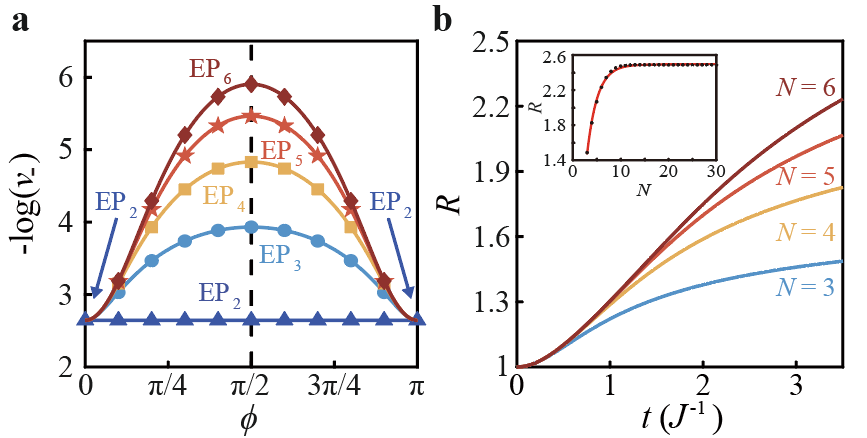}
		\caption{\textbf{Entanglement enhancement by higher-order EPs.}
			\textbf{a} Logarithmic negativity $-\log\nu_-$ of $N$-mode BKC model as a function of hopping phase $\phi$ for $N = 2$ (blue), $N = 3$ (light blue), $N = 4$ (yellow), $N = 5$ (orange), and $N = 6$ (red). Here, we set $Jt=3.5$ and any point on the solid line corresponds to 2-fold EP$_N$s except for $\phi=0,\pi$ (X-fold EP$_2$s instead).  \textbf{b} The time evolution of optimized entanglement enhancement ratio $R(t)=\log[\nu_-(\pi/2,t)]/\log[\nu_-(0,t)]$ with different lengths of $N=3$ (light blue), $N=4$ (yellow), $N=5$ (orange) and $N=6$ (red). The inset shows the ratio R as a function of $N$ ($2\sim30$) with fixed time, i.e., $Jt=3.5$, which can be fitted as $R\approx-4.11 e^{-0.4633 N}+2.493$, revealing a trend of nearly 2.5 times enhancement by highest-order EPs in thermodynamic limits. The other parameters are fixed at $g/J=1$ and $\eta=0$.
		}
		\label{fig3}
	\end{figure}
	\noindent
	Beyond second-order EPs, we demonstrate that $N$-mode chains have the potential to exhibit higher-order EPs, which can induce stronger entanglement compared to the case of second-order EPs. In previous work, higher-order EPs can be generated by the evolution matrices of system operator higher-order moments\upcite{Arkhipov2021}. Here, we use an alternative method to achieve higher-order EPs of bosonic BdG matrix by introducing a non-integer-$\pi$ hopping phase, efficiently constructing a dynamical matrix with a higher-dimensional Jordan block (see details in Supplementary Note 2).
	Considering an $N$-mode BKC, 2-fold highest-order EPs (EP$_N$s) emerge in the dynamical spectrum when the phase $\phi$ of the BS terms satisfies $\phi\neq m\pi$ ($m \in \mathbb{Z}$) with $g/J=1$ (For $g/J\neq1$, the system is not operated at EP regardless of the phase $\phi$. See details in Supplementary Note 2). Conversely, $X$-fold EP$_2$s arise at $\phi = m\pi$ as discussed before.  
	Under the same parameter condition, we observe that the degree of multimode entanglement is significantly higher when the system operates at EP$_N$s compared to EP$_2$s. As illustrated in Fig.~\ref{fig3}(a), the (1$|$N-1)-bipartite entanglement degree increases to its maximum as the hopping phase varies from $0$ to $\pi/2$.  
	The corresponding $\nu_-(\phi,t)$ in the $N$-mode system is expressed as 
	\begin{equation}
		\nu_-(\phi,t)=\sqrt{\xi_N(\phi,t)-\sqrt{\xi_N^2(\phi,t)-1}},
	\end{equation}
	with
	$\xi_N(\phi,t)=1+\sum^{N-1}_{j=1}c_j J^{2j}t^{2j}\sin^{2(j-1)}(\phi)$, where the coefficient sequence $\{c_j\}$ remains consistent across different $N$ (see details in Supplementary Note 3). 
	Obviously, a non-integer-$\pi$ phase induces higher-order terms in $\xi_N$ leading to enhanced entanglement and thereby directly demonstrating the connection between the order of EPs and entanglement dynamics.
	Besides, We can find that $-\log(\nu_-)$ increases monotonically with the phase distance from $\phi=0$ to $\pi/2$.  
	
	Moreover, the degree of multimode entanglement in the highest-order EPs accumulates with increasing time, as shown in Fig.~\ref{fig3}(b). The improvement of the optimized multimode entanglement (the maximum at $\phi=\pi/2$) induced by EP$_N$s for the $N$-mode BKC can be quantified by the ratio $R(t) = \log[\nu_-(\pi/2,t)] / \log[\nu_-(0,t)]$, which exhibits exponential growth with the order of EPs [see Fig.~\ref{fig3}(b) inset]. 
	As $N$ increases, we find that multimode entanglement induced by the highest-order EPs can reach up to $\sim$2.5 times that of the EP$_2$s case at the fixed time $Jt=3.5$. The saturation feature observed here is caused by the coefficients $c_j$ which decrease by orders of magnitude as $j$ increases. More specifically, although the highest power $(Jt)^{2(N-1)}$ grows rapidly with $N$, its contribution is suppressed by the small prefactor $c_j$ (See details in Supplementary Note 3). So, as $N$ increases, the additional gain from higher-order terms diminished for a fixed $Jt$, leading to the observed saturation of $R$.

	\begin{figure}
		\centering
		\includegraphics[width=8cm]{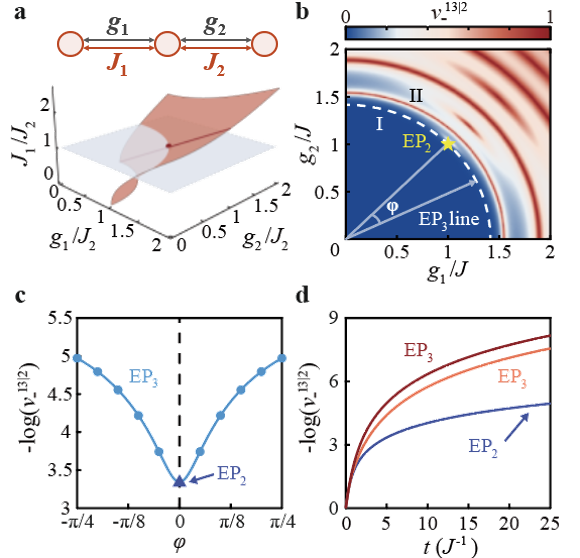}
		\caption{\textbf{Exceptional points and entanglement dynamics in the nonuniform BKC.}
			\textbf{a} Schematic of a three-mode system and the corresponding conditions for EPs in parameter space. The light red surface represents the 2-fold third-order ESs, where each point corresponds to 2-fold EP$_3$s, except for the EAs (red line) consisting of 2-fold EP$_2$s. The parameter space is divided into two regions: eigenvalues are purely imaginary above the ESs, while they are purely real below the ESs, with the exception of two eigenvalues always being 0. The light blue cut at $J_1/J_2 =1$ is shown as a visual guide for the data presented in panel \textbf{b}. 
			\textbf{b} The entanglement witness $\nu_-^{13|2}$ at $Jt = 5$ versus interaction strengths, along the blue cut in panel \textbf{a}. 
			In region I, entanglement evolves exponentially, while in region II, it oscillates. The white dashed line, marking the intersection of the light red surface and the light blue plane, indicates the positions of EPs, with $\varphi$ representing the angle between EP$_3$s and EP$_2$s.  
			\textbf{c} Entanglement witness $-\log(\nu_-^{13|2})$ as a function of $\varphi$ along the white dashed line in \textbf{b}. The degree of entanglement increases as $\varphi$ moves further from EP$_2$s.  
			\textbf{d} Time evolution of $-\log(\nu_-^{13|2})$  at $\varphi = 0$ (blue line), $\varphi = \pi/8$ (orange line), and $\varphi = \pi/4$ (red line), with other parameters being the same as in \textbf{c}.  
		}
		\label{fig4}
	\end{figure}

	Besides adding a non-integer-$\pi$ hopping phase, tuning BS (TMS) interaction strengths being unequal, i.e., $g_{i}\neq g_{j}$ ($J_{i}\neq J_{j}$), can also be applied to construct the highest-order EPs. Taking a 3-mode BKC as an example [Fig.~\ref{fig4}(a)], the condition for generating EPs  requires $g_1^2 + g_2^2 = J_1^2 + J_2^2$  (For $g_1^2 + g_2^2 \neq J_1^2 + J_2^2$, the system is not operated at EPs. See details in Supplementary Note 2). Unlike the emergence of third-order EPs induced by the hopping phase, 2-fold third-order exceptional surfaces (ESs) can be observed with zero hopping phases.  
	The degenerate ESs divide the parameter space into two distinct regions: one where all eigenvalues are purely imaginary (above the ESs) and another where they are purely real (below the ESs).  
	Additionally, the parameter space of the degenerate ESs includes a boundary line (red line), corresponding to the case where $g_1 = J_1$ and $g_2 = J_2$. This condition generates the second-order exceptional arcs (EAs)\upcite{Tang2020}.  
	
	Figure~\ref{fig4}(b) illustrates the entanglement behavior under varying hopping strengths [along the light blue cut in Fig.~\ref{fig4}(a), $J_{1}=J_{2}=J$].  
	By calculating $\nu_-^{13|2}$ to measure the bipartite entanglement between mode $a_2$ and modes $a_{1,3}$, the degree of entanglement displays two distinct behaviors, separated by the EPs (white dashed line). When $g_1^2 + g_2^2 < 2J^2$, an exponentially decaying trend (type I) is observed, while oscillations (type II) occur for $g_1^2 + g_2^2 > 2J^2$.  
	This behavior aligns with the entanglement dynamics governed by EPs in the BKC with identical interaction strengths, as discussed earlier. Similar patterns are observed for the other two bipartite entanglements, $\nu_-^{12|3}$ and $\nu_-^{23|1}$, further indicating the presence of genuine tripartite entanglement.
	The entanglement gain induced by the highest-order EPs can be quantified by varying the interaction strengths along the white dashed line, as described by the equation: $\nu_-^{13|2}(\varphi,t) = (\xi_{\mathrm{n-id}}(\varphi,t) - \sqrt{\xi_{\mathrm{n-id}}^2(\varphi,t) - 1})^{1/2}$, with
	$
	\xi_{\mathrm{n-id}}(\varphi,t) = 32J^4t^4\sin^2\varphi + 16J^2t^2 + 1.
	$
	Here, the parameter $\varphi=\pi/4 - \arctan(g_{2}/g_{1})$ represents the angle between EP$_3$s and the EP$_2$s (see details in Supplementary Note 3). 
	The degree of entanglement as a function of $\varphi$ and $t$ is shown in Fig.~\ref{fig4}(c) and (d), which exhibits the stronger entanglement induced by higher-order EPs.

	\section*{Discussion}
	\noindent
	Non-Hermitian physics has garnered considerable attention in recent years, driving advancements in sensing and wave manipulation. As quantum techniques develop, it has become feasible to simulate non-Hermitian dynamics within quantum systems, attracting both theoretical and experimental interest in how to harness quantum resources under non-Hermitian conditions effectively. 
	By constructing a non-Hermitian BdG framework to understand multimode entanglement dynamics, our work provides a foundational and distinct mechanism to utilize entanglement resources efficiently. We present two examples to demonstrate its potential applications.
	In quantum metrology, for a parameter $\chi$ encoded in an initial state, the metrological power of the corresponding final quantum state can be quantified by its quantum Fisher information (QFI) $Q_\chi$, which sets a bound on the precision $(\Delta \chi)^2$ to estimate $\chi$, i.e., $(\Delta \chi)^2 \ge 1/(\nu Q_\chi)$ with $\nu$ being the number of independent measurements. Thus, a larger QFI indicates greater metrological power. We can find that the QFI $Q_\chi$ in our three-mode BKC is significantly enhanced in the parameter regimes where all nonzero eigenvalues are purely imaginary, while it remains relatively low in the purely real regimes. This demonstrates that distinct behaviors of QFI on the two sides of EP$_2$s can be observed and the states in purely imaginary region can provide a higher accuracy for parameter estimation.
	Another application is to detect multimode entanglement under noise. 
	For a fixed evolution time, we can find the entanglement in our three-mode BKC quickly vanishes as the thermal noise increases in the purely real spectral region. In contrast, in the region where all nonzero eigenvalues are purely imaginary, the entanglement only diminishes slightly and remains detectable, revealing a more resilient parameter regime for implementing and observing multimode entanglement under noise (see detailed calculations in Supplementary Note 4).

	To discuss the possible implementation of our model, we note that the rapid progress in quantum technologies---particularly in bosonic platforms---offers promising avenues for realizing both mode-squeezing and mode-exchange interactions. Optical and mechanical systems are especially well-suited for this purpose, as their photonic and mechanical degrees of freedom naturally support continuous-variable quantum information processing and bosonic quantum simulations.
	For instance, integrated optical chips have recently emerged as powerful platforms capable of implementing frequency-mode exchange and pair-squeezing interactions among quantum modes\upcite{jia2025, Wang2025}.
	In optomechanics\upcite{Del2022, Wanjura2023, Slim2024}, exchange (squeezing) couplings can be engineered by modulating the intensity of the drive laser using multiple harmonic tones with frequencies that approximate the frequency differences (sums) of mechanical modes. Moreover,  two- and three-mode hopping and single-mode squeezing interactions were demonstrated in a gigahertz-frequency multimode mechanical resonator coupled to a superconducting qubit\upcite{vonLupke24, Marti2024}. By controlling the qubit's driving frequency, these exchange and squeezing couplings can also be extended to a large-scale multimode setting.
	Additionally, superconducting quantum circuits also offer a natural environment for realizing hopping and pairing terms in the synthetic dimensions of a multimode superconducting parametric cavity\upcite{Busnaina2024}.

	In summary, our work studies the entanglement behavior driven by EPs in multimode bosonic models. By mapping the Heisenberg equations to the non-Hermitian BdG framework, we uncover a variety of EPs, from second-order to highest-order, separating distinct spectral regimes characterized by purely imaginary, purely real, or mixed eigenvalue spectra. These EPs directly determine nonequilibrium entanglement transitions, giving rise to exponential, oscillatory, or hybrid entanglement dynamics.  
	Remarkably, by comparing to the multimode entanglement achievable in the EP$_2$s case, we demonstrate that the highest-order EPs, obtained through non-integer-$\pi$ hopping phases or nonuniform interaction strengths, can be designed to enhance the entanglement degree when operated optimally. 
	Our findings provide new insights into the relationship between non-Hermitian features and quantum entanglement, with the potential applications for further engineering quantum correlations as well as enhancing sensitivity in photonic\upcite{jia2025,Wang2025}, optomechanical\upcite{Del2022,Wanjura2023,Slim2024}, and mechanical platforms\upcite{Bild2023,vonLupke24,Marti2024,Yang2024}.

	Looking forward, several intriguing questions remain open. 
	For example, (i) while we provide a connection between second-order quadrature correlations (covariance matrix) and the dynamical matrix, it is unclear whether a deeper connection can be established between the dynamical matrix and higher-order quadrature correlations. This would be valuable, as it may offer a path to better understand multimode non-Gaussian entanglement dynamics~\upcite{tian2025b} (i.e., entanglement beyond second-order quadrature correlations), which is considered as a necessary resource for quantum computation and metrology~\upcite{niset2009,tian2022}.
	(ii) The introduction of the squeezing terms may also induce rich topological phases\upcite{McDonald2018,Busnaina2024,Slim2024}. Given the strong connection between the topological phases and the dynamics of first-order moments\upcite{Metelmann2015,Porras2019,Wanjura2020,Tian2023,Gomez-Leon2023,Wanjura2021}, the investigations of the connection with the CM could provide valuable insights into a deeper understanding of quantum correlations in a topological framework.
	(iii) It would also be interesting to extend dissipation-free bosonic chains to Markovian\upcite{Wu2019,Naghiloo2019,Minganti2019,Arkhipov2021} and non-Markovian\upcite{Lin2025} regimes and investigate the connection between the non-Hermiticity  and quantum correlations in open quantum systems.

	\section{Methods}
	\subsection{The positive partial transpose criterion}
	\noindent
	Here, we introduce the entanglement criterion we used to describe the multimode entanglement in our model. For Gaussian systems, the positive partial transpose (PPT) criterion is widely applied to measure the bipartite entanglement\upcite{Simon2000,Adesso2007}. Specifically, considering a $n_A\times m_B$ bipartite Gaussian state $\rho_{AB}$, we can obtain its CM $\sigma_{AB}$ with matrix elements $\sigma_{ij}=\langle \hat{\beta}_i\hat{\beta}_j+\hat{\beta}_j\hat{\beta}_i\rangle-2\langle \hat{\beta}_i\rangle\langle \hat{\beta}_j\rangle$, where we define vector $\pmb{\beta}=(\hat{X}_1^A,\hat{P}_1^A,\ldots,\hat{X}_n^A,\hat{P}_n^A,\hat{X}_1^B,\hat{P}_1^B,\ldots,\hat{X}_m^B,\hat{P}_m^B)$. As a bona fide CM, it should satisfy 
	\begin{equation}
		\sigma_{AB}+i\Omega\ge0,
	\end{equation}
	or equivalently
	\begin{equation}
		\nu_i^{AB}\ge1 \;(\forall i =1,\ldots,n_A+m_B),
	\end{equation}
	where $\nu_i^{AB}$ is the symplectic eigenvalue of $\sigma_{AB}$ and $\Omega$ is the symplectic form
	\begin{equation}
		\Omega=\bigoplus_{k=1}^N\omega,\quad \omega =
		\begin{bmatrix}
			0&1\\-1&0
		\end{bmatrix}.
	\end{equation}
	Correspondingly, the PCM $\tilde{\sigma}_{AB} = \theta_{AB}\sigma_{AB}\theta_{AB}$ with
	\begin{equation}
		\theta_{AB} = \mathrm{diag}\{\underbrace{1,1,\ldots,1,1}_{2n_A},\underbrace{1,-1,\ldots,1,-1}_{2m_B}\},
	\end{equation}
	which means mapping $\hat{P}_B$ to $-\hat{P}_B$. PPT criterion indicates that if the new matrix $\tilde{\sigma}_{AB}$ isn't a bona fide CM, i.e.
	\begin{equation}
		\tilde{\sigma}_{AB}+i\Omega<0, 
	\end{equation}
	or equivalently
	\begin{equation}
		\tilde{\nu}_i^{AB}<1 \;(\exists i \in \{1,\ldots,n_A+m_B\}),
	\end{equation}
	the subsystem $S_A$ composed of $n_A$ modes is entangled with the subsystem $S_B$ composed of $n_B$ modes. To be clear, the PPT criterion is the sufficient and necessary condition for all ($n_A\times1$)-mode Gaussian states but not for generic cases. However, it still can be used to measure the entanglement in $n_A\times m_B$ bipartite Gaussian state once $\tilde{\sigma}_{AB}$ isn't a bona fide CM.
	Moreover, we can used the logarithmic negativity $E_{\mathcal{N}}$ to quantify the violation of the PPT criterion, i.e.
	\begin{equation}
		E_{\mathcal{N}}=\left\{
		\begin{aligned}
			&-\sum_{k} \log\tilde{\nu}_k,&\mathrm{for}&\quad k:\tilde{\nu}_k<1\\
			&0,&\mathrm{if}& \quad \tilde{\nu}_i\ge1\forall i.
		\end{aligned}\right.
	\end{equation}
	Namely, the logarithmic negativity $E_{\mathcal{N}}$ is a good choice to measure entanglement. By the way, there is a lemma that at most $N_{\mathrm{min}}\equiv\mathrm{min}\{n_A,m_B\}$ symplectic eigenvalues $\tilde{\nu}_k$ of $\tilde{\sigma}_{AB}$ can violate the PPT inequality for a $n_A\times m_B$ bipartite Gaussian state. Especially, for all $(n_A\times 1)$-mode Gaussian states, we can directly use $\nu_-\equiv\min\{\tilde{\nu}_i\}$ to measure entanglement as we apply to measure the entanglement between one mode and the rest of the system, denoted as $(1|N-1)$, in the main text. When $\nu_-<1$, the system has entanglement. Otherwise, it's separable. Furthermore, the closer $\nu_-$ to zero, the greater the entanglement.

	\section*{Data availability}
	\noindent
	The data that support the findings of this study are available from the corresponding author upon reasonable request.
	
	\section*{Acknowledgements}
	\noindent
    This work is supported by Quantum Science and Technology-National Science and Technology Major Project (Grant No. 2024ZD0302401), National Natural Science Foundation of China (Grants No. 12125402, No. 12534016, and No. 12474354), Beijing Natural Science Foundation (Grant No. Z240007), and the Fundamental Research Funds for the Central Universities.
	
	\section*{Author contributions}
	\noindent
	Chenghe Yu carried out the calculations. Mingsheng Tian, Xinyao Huang and Chenghe Yu analyzed the results and wrote the paper. Xinyao Huang conceived the idea.  Qiongyi He supervised the project. All authors contributed to discussions and writing of the manuscript.
	
	\section*{Competing interests}
	\noindent
	The authors declare no competing interests.
	
	\section*{Additional information}
	\noindent
	\textbf{Supplementary information} The online version contains supplementary material available at https://XXXXX.
	
	\bigskip
	
	\noindent
	\textbf{Correspondence} and requests for materials should be addressed to Xinyao Huang or Qiongyi He.

	
	

	\renewcommand\bibnumfmt[1]{#1.}
	\bibliography{refs_light.bib}
	\bibliographystyle{naturemag}
\end{document}